\documentclass[iop,apj,tighten]{emulateapj}
\usepackage[breaklinks,colorlinks,urlcolor=blue,citecolor=blue,linkcolor=blue]{hyperref}
\usepackage{graphicx}
\usepackage{bm}
\usepackage{amsmath,amssymb}
\usepackage{appendix}
\usepackage{color}
\newcommand{\Msun}{{\rm M}_\odot}

\shorttitle{Eccentricity distribution in AGNs}
\shortauthors{Tagawa et al.}

\begin{document}

\title{Eccentric Black Hole Mergers in Active Galactic Nuclei}
  
\author{Hiromichi Tagawa\altaffilmark{1}, 
Bence Kocsis\altaffilmark{2}, 
Zolt{\'a}n Haiman\altaffilmark{3}, 
Imre Bartos\altaffilmark{4}, 
Kazuyuki Omukai\altaffilmark{1}, 
and 
Johan Samsing\altaffilmark{5}
}
\affil{\altaffilmark{1}Astronomical Institute, Graduate School of Science, Tohoku University, Aoba, Sendai 980-8578, Japan\\
\altaffilmark{2}{Rudolf Peierls Centre for Theoretical Physics, Clarendon Laboratory, Parks Road, Oxford, OX1 3PU, UK}\\
\altaffilmark{3}Department of Astronomy, Columbia University, 550 W. 120th St., New York, NY, 10027, USA\\
\altaffilmark{4}{Department of Physics, University of Florida, PO Box 118440, Gainesville, FL 32611, USA}\\
\altaffilmark{5}Niels Bohr International Academy, The Niels Bohr Institute, Blegdamsvej 17, 2100 Copenhagen, Denmark
}
\email{E-mail: htagawa@astr.tohoku.ac.jp}

\begin{abstract} 
The astrophysical origin of gravitational wave (GW) transients is a timely open question in the wake of discoveries by LIGO/Virgo. In active galactic nuclei (AGNs), binaries form and evolve efficiently by interaction with a dense population of stars and the gaseous AGN disk. 
Previous studies have shown that stellar-mass black hole (BH) mergers in such environments can explain the merger rate and the number of suspected hierarchical mergers observed by LIGO/Virgo. 
The binary eccentricity distribution can provide further information to distinguish between astrophysical models. 
Here we derive the eccentricity distribution of BH mergers in AGN disks. 
We find that eccentricity is mainly due to binary-single (BS) interactions, which lead to
most BH mergers in AGN disks having a significant eccentricity at $0.01\,\mathrm{Hz}$, detectable by LISA.  
If BS interactions occur in isotropic-3D directions, then  $8$--$30\%$ of the mergers in AGN disks will have eccentricities at $10\,\mathrm{Hz}$ above  $e_{10\,\rm Hz}\gtrsim 0.03$, detectable by LIGO/Virgo/KAGRA, while $5$--$17\%$ of mergers have 
$e_{10\,\rm Hz}\geq 0.3$. 
On the other hand, if BS interactions are confined to the AGN-disk plane 
due to torques from the disk, 
with 1-20 intermediate binary states during each interaction, or if BHs can migrate to $\lesssim10^{-3}\,\mathrm{pc}$ from the central supermassive black hole, 
then $10$--$70\%$ of the mergers will be highly eccentric ($e_{10\,\rm Hz} \geq 0.3$), consistent with the possible high eccentricity in GW190521. 
\end{abstract}

\section{Introduction}

Recent detections of gravitational waves (GWs) have shown evidence for a high black hole (BH) merger rate in the Universe \citep{TheLIGO18}. 
However, the proposed astrophysical pathways to merger remain debated. 

Measuring the binary eccentricity ($e$) is useful to distinguish between possible astrophysical pathways to merger. 
The feasibility to measure $e$ has increased tremendously due to the improvement of the detectors and GW data analysis methods \citep[e.g.][]{Nishizawa16,Lower18,Abbott19_Ecc,Romero-Shaw19}. 
At design sensitivity, the LIGO, Virgo, and KAGRA detectors 
may detect eccentricities at $10\,\mathrm{Hz}$ above $e_\mathrm{10Hz} \gtrsim 0.03$ \citep{Lower18,Gondan_Kocsis19,Romero-Shaw19}, while LISA will detect $e$ if above $10^{-3}$--$10^{-2}$ at $0.01\,\mathrm{Hz}$ \citep{Nishizawa16}. 
Eccentricity varies greatly among different merger channels \citep{Antonini12,Breivik16,Silsbee17,Antonini17,ArcaSedda18,2018MNRAS.481.5445S, Rodriguez18bR,Fragione_Bromberg2019,Fragione+2019,Zevin19,Martinez+20}, but only binaries formed through GW capture (GWC) are generally expected to produce mergers with $e_\mathrm{10Hz}\gtrsim 0.1$ \citep{2006ApJ...648..411K,OLeary09,Gondan18a,Samsing18b,Rasskazov19,Gondan20_GW190521}. 

Recently, the merger of two unusually heavy BHs, GW190521, was reported  \citep{LIGO20_GW190521,LIGO20_GW190521_astro}. 
Their high masses ($85^{+21}_{-21}\,\Msun$ and $66^{+17}_{-18}\,\Msun$) and high spins ($0.69^{+0.27}_{-0.62}$ and $0.73^{+0.24}_{-0.64}$) indicate 
that the merging BHs themselves could have been the remnants of earlier BH mergers, which increased their masses beyond the $\sim56$\,M$_\odot$ limit due to pulsational pair-instability \citep{Farmer19} 
and their spins beyond the small natal values expected in stellar evolutionary models of angular momentum transfer \citep{Fuller19_massive}. 
Thus, detection of GW190521 (and also other events: GW170729, GW170817A, GW190412 and GW190814; \citealt{Zackay19_GW170817A,LIGO20_GW190412,LIGO20_GW190814}) suggests 
that hierarchical mergers could be frequent among compact objects. This is consistent with the scenario of mergers in AGN disks (e.g. \citealt{Yang19b_prl,Yang20_gap}; \citealt{Tagawa20b_spin},  hereafter Paper~II), 
in which binaries are efficiently hardened by interaction with the surrounding gas  \citep[e.g.][]{Bartos17,Stone17,McKernan17} and with the dense populations of stars and compact objects \citep[][hereafter Paper~I]{Tagawa19}. 
A possible electromagnetic counterpart would further support this scenario \citep{2019ApJ...884L..50M,Graham20}. 

Interestingly, \citet{Gayathri20_GW190521} 
found that GW190521 prefers a high eccentricity of $e_\mathrm{10Hz}\sim 0.7$, 
along with high spin-precession \citep[see also][]{Romero-Shaw20}. 
A high eccentricity places additional constraints on possible
astrophysical models of this source
\citep[e.g.][]{Zevin19,Rodriguez18bR,Rasskazov19}. 

\citet{Samsing20} 
recently showed that highly-eccentric mergers are common in AGN disks by assuming that binary-single (BS) interactions of BHs are confined to a plane. That study assumed constant values for the frequency of BS interactions, initial separation, and relative velocity of a third body based on Paper~I. 
In this {\em Letter}, we investigate the distribution of the binary eccentricity for mergers in AGN disks by performing one dimensional (1D) $N$-body simulations combined with semi-analytical prescriptions, which enable us to follow binaries considering such effects as eccentricity evolution due to BS interactions, type I/II torques exerted by circumbinary disks, and GW radiation ($\S\,\ref{sec:evolution_binary}$).  We also augment the model used in Papers~I/II to include GW capture in single-single (SS) and BS encounters.

\section{Method}

\begin{table*}
\begin{center}
\caption{Fiducial values of our model parameters. 
}
\label{table:parameter_model}
\hspace{-5mm}
\begin{tabular}{c|c}
\hline 
Parameter & Fiducial value \\
\hline\hline
Spatial directions in which BS interactions occur&  isotropic in 3D\\\hline
Number of temporary binary BHs formed during a BS interaction&  $N_\mathrm{int}=20$\\\hline
Initial BH spin magnitude & $|{\bm a}|=0$\\\hline
Angular momentum directions of circum-BH disks & $\hat{{\bm J}}_\mathrm{CBHD}=\hat{{\bm J}}_\mathrm{AGN}$ for single BHs,\\&
$\hat{{\bm J}}_\mathrm{CBHD}=\hat{{\bm J}}_\mathrm{bin}$ for BHs in binaries\\\hline
Ratio of viscosity responsible for warp propagation\\over that for transferring angular momentum 
& $\nu_2/\nu_1=10$ \\\hline
Alignment efficiency of the binary orbital angular momentum
due to gas capture 
& $f_\mathrm{rot}=1$
\\
(Eq.~14 in Paper~II)&\\
\hline 
Mass of the central SMBH & $M_\mathrm{SMBH}=4\times 10^6\,\Msun$ \\\hline
Gas accretion rate at the outer radius of the simulation ($5\,\mathrm{pc}$)
& ${\dot M}_\mathrm{out}=0.1\,{\dot M}_\mathrm{Edd}$ with $\eta=0.1$\\\hline
Fraction of pre-existing binaries & $f_\mathrm{pre}=0.15$ \\\hline
Power-law exponent for the initial density profile for BHs & $\gamma_{\rho}=0$ \\\hline
Initial velocity anisotropy parameter\\such that $\beta_\mathrm{v}v_\mathrm{kep}(r)$ is the BH velocity dispersion  
& $\beta_\mathrm{v}=0.2$ \\\hline
Efficiency of angular momentum transport in the $\alpha$-disk & $\alpha_\mathrm{SS}=0.1$ \\\hline
Stellar mass within 3 pc &$M_\mathrm{star,3pc}=10^7\,\Msun$\\\hline 
Stellar initial mass function slope & $\delta_\mathrm{IMF}=2.35$\\
(relation between the stellar and BH masses are in Eq.~3 of Paper~I) &\\
\hline
Angular momentum transfer parameter in the outer star forming regions 
&$m_\mathrm{AM}=0.15$\\
(Eq.~C8 in \citealt{Thompson05}) &\\
\hline
Accretion rate in Eddington units onto\\stellar-mass BHs with a radiative efficiency $\eta=0.1$
&$\Gamma_\mathrm{Edd,cir}=1$\\\hline
Numerical time-step parameter &$\eta_t=0.1$\\\hline
Number of radial cells storing physical quantities &$N_\mathrm{cell}=120$\\\hline
Maximum and minimum $r$ for the initial BH distribution&  $r_\mathrm{in,BH}=10^{-4}$ pc, $r_\mathrm{out,BH}=3$ pc \\\hline
Initial number of BHs within 3 pc &$N_\mathrm{BH,ini}=2\times 10^4$ \\\hline
\end{tabular}
\end{center}
\end{table*}

Our model is based on that in Paper~I and partially in Paper~II as specified below. 
The new ingredients are described in some detail in $\S\,\ref{sec:GW_SS}$, $\S\,\ref{sec:dynamical_effects}$ and $\S\,\ref{sec:ecc_ev}$. 

\subsection{Components}

We suppose that there is a supermassive BH (SMBH) at the center of a galaxy 
and it has a gaseous accretion disk (hereafter "AGN disk") and a spherically distributed stellar cluster (hereafter "spherical component"). 
We follow the evolution of the BH system consisting of 
a cluster of BHs around the AGN disk (called the flattened component due to its spatial distribution, which is assumed to be caused by vector resonant relaxation; \citealt[][]{Szolgyen18}), 
and stars and BHs captured inside the AGN disk
due to the gaseous torque (called the disk stellar/BH components). 
See Fig.~1 in Paper~I for an illustration of these components. 

\subsection{Binary formation and disruption}

\label{sec:binary_formation_overview}
Some BHs in the flattened component 
are in binaries from the outset (called {\it pre-existing binaries}). 
In the AGN disk, binaries form due to gas dynamical friction during two-body encounters
(dubbed the {\it gas-capture binary formation}). 
Binaries also form due to dynamical interactions during three-body encounters 
({\it dynamical binary formation}).
At their formation, a thermal distribution $f(e)=2e$ is assumed, 
while $e=0$ was assumed in Papers~I/II. 
In addition to those mechanisms already included in Paper~I, 
we here consider the binaries formed by the GWC
mechanism \citep[e.g.][]{OLeary09, 2020PhRvD.101l3010S}, which are relevant for highly eccentric mergers, as described below in $\S\,\ref{sec:GW_SS}$. 
Binaries are disrupted by soft-BS interactions, in which the binaries become softer 
\citep[e.g.][]{Heggie75}. 

\subsubsection{Binary formation by the GW capture}
\label{sec:GW_SS}
We treat binary formation by GWC due to SS encounters (SS-GWC) and BS interactions (BS-GWC) separately
as described below. 

In an SS encounter event, 
a binary can form if the two bodies approach close enough that the energy radiated by GWs ($\Delta E_\mathrm{GW}$) exceeds 
their kinetic energy, 
$E_\mathrm{SS}\approx \frac12\mu v_{12}^2$, where $\mu$ is the reduced mass, and $v_{12}$ the relative velocity. 
$\Delta E_\mathrm{GW}$ is approximately given as 
\begin{equation}
\label{eq:de_gw}
    \Delta E_\mathrm{GW}\approx \frac{85\pi}{12\sqrt{2}}
    \frac{G^{7/2}\eta^2 m_\mathrm{tot}^{9/2}}{c^5r_\mathrm{p}^{7/2}},
\end{equation}
\citep[e.g.][]{OLeary09}, 
where $G$ is the gravitational constant, $c$ the light speed, 
$m_\mathrm{tot}$ the total mass of the two bodies, $\eta$ 
the symmetric mass ratio, 
and the pericenter distance
\begin{equation}
\label{eq:rp}
r_\mathrm{p} = \left(\sqrt{\frac{1}{b^2}+\frac{G^2m_\mathrm{tot}^2}{b^4 v_\mathrm{12}^4}} + \frac{Gm_\mathrm{tot}}{b^2 v_{12}^2}\right)^{-1} 
\end{equation}
where $b$ is the impact parameter of the encounter. 
By equating $\Delta E_\mathrm{GW}$ in equation (\ref{eq:de_gw}) and $E_\mathrm{SS}$, 
the maximum pericenter distance for the GWC is 
\begin{equation}
\label{eq:rp_max}
r_\mathrm{p,max}=\left(\frac{85\pi \eta c^2}{6\sqrt{2} v_{12}^2}\right)^{2/7} 
\frac{Gm_\mathrm{tot}}{c^2}. 
\end{equation}
Note that the gas has negligible impact on the dynamics during the GWC owing to very short timescale for the capture ($\sim \mathrm{hr}$,  \citealt{OLeary09}). 
Here we define the maximum impact parameter $b_\mathrm{max}$ at which $r_\mathrm{p}=r_\mathrm{p,max}$. We assume that the cross section of SS-GWC $\sigma_\mathrm{SS}$ is 
approximately given as 
$\sigma_\mathrm{SS} = r_\mathrm{SS} z_\mathrm{SS}$, 
where $r_\mathrm{SS}=\mathrm{min}[b_\mathrm{max},r_{\mathrm{Hill}}]$, $z_\mathrm{SS}=\mathrm{min}[r_\mathrm{SS},h_\mathrm{c}]$, $r_{\mathrm{Hill}}=r(m_\mathrm{BH}/3M_\mathrm{SMBH})^{1/3}$ is the Hill radius with respect to the SMBH, $m_\mathrm{BH}$ is the mass of the BH, 
and $h_\mathrm{c}$ is the average orbital height of background objects at the radial location of the BH (see Paper~I).

For each object, the timescale to undergo the SS-GWC is 
$t_\mathrm{GSS}=1/(n_\mathrm{c}\sigma_\mathrm{SS}v_{\mathrm{rel}})$, where $n_\mathrm{c}$ is the number density of background objects, and $v_{\mathrm{rel}}$ the typical relative velocity between the BH and background objects (Paper~I). 
Here we substitute $v_{\mathrm{rel}}=v_{12}$ 
in Eqs.~\eqref{eq:de_gw}--\eqref{eq:rp}. 
We set the probability of binary formation by this mechanism during the timestep $\Delta t$ to $P_\mathrm{GSS}=\Delta t/t_\mathrm{GSS}$. 
When the SS-GWC binaries form, 
we choose $b$ so that $b^2$ is uniformly distributed  between 0 and $b_\mathrm{max}^2$.
The semi-major axis and $e$ of the newly formed binary are 
$s=Gm_\mathrm{tot}^2 \eta/(2|\Delta E_\mathrm{GW}|-2 E_\mathrm{SS})$, 
and 
$e=1-r_\mathrm{p}/s$, respectively. 
We only consider binary formation through SS-GWC between disk BHs, 
as the probability of interactions 
is much lower 
for spherical component objects 
due to their large $v_{\mathrm{rel}}$
($P_\mathrm{SS}\propto v_{\mathrm{rel}}^{-11/7}$ 
for $b_\mathrm{max}< h_\mathrm{c}, r_{\mathrm{Hill}}$), 
and due to the large physical sizes of stars.
For binaries formed in this process, we assume that the orbital angular momentum directions are aligned (or half anti-aligned) 
with the angular momentum direction of the AGN disk. 

GW-capture binaries can also form during BS interactions \citep[e.g.][]{Samsing14} 
if one body is captured due to strong GW emission. 
Before the interaction, the velocities among BHs in the flattened component are likely distributed almost randomly in 3D space, while for interactions among the disk BHs the three-body velocities may be constrained in the AGN-disk plane due to the disk gas drag.  
We here study the 3D and 2D hard-BS interactions separately. 
In the case of the 3D hard-BS interactions, 
the probability of GW-capture binary formation is roughly given by 
\begin{equation}
\label{eq:p_cap_iso}
    P_\mathrm{GBS}\simeq N_\mathrm{int} \frac{2 r_\mathrm{p,max} }{s}
\end{equation}
\citep{Samsing18b}, 
where $N_\mathrm{int}$ is the average number of quasi-stable states for 
temporary binary BHs formed during a BS interaction, 
where $N_\mathrm{int}\sim 20$ is numerically verified for both 2D and 3D interactions in 
equal-mass cases (\citealt{Samsing14}; \citealt{Samsing20}). 
The GWC probability is calculated as 
\begin{equation}
\label{eq:p_cap_disk}
    P_\mathrm{GBS}=1-\left(1- P_\mathrm{GBS,1}\right)^{N_\mathrm{int}}
\end{equation}
where
\begin{equation}
\label{eq:p_cap1n}
    P_\mathrm{GBS,1}=\frac{\int_{e_\mathrm{cr}}^1 f(e)de }{\int_0^1f(e)de}
\end{equation}
is the probability that GWC occurs in one interaction, 
$e_\mathrm{cr}=1-r_\mathrm{p,max}/s$ is the critical eccentricity above which the GWC binaries form due to GW emission, 
and $f(e)$ is either $f_{\rm 3D}=2e$ \citep[e.g.][]{Heggie75} or $f_{\rm 2D}=e/\sqrt{1-e^2}$ \citep{Valtonen06} for isotropic-3D and 2D interactions, respectively.

When the BS interaction ends in a GW-capture, the semi-major axis $s$ does not
change from 
that before the BS interaction, 
we determine $r_\mathrm{p}=s(1-e)$ 
by drawing $e$ from $f(e)$. 
This assumption is due to energy conservation, since in this case the third body does not receive a strong velocity kick and it is not ejected. 
Capture occurs only if $r_{\rm p}<r_\mathrm{p,max}$ (Eq.~\ref{eq:rp_max}). 
We assume that $v_{12}=\sqrt{Gm_\mathrm{tot}/s}$, i.e. the intrinsic orbital velocity of the binary. 
For simplicity, we also assume that the BS-GWC results in binary formation among the most and second-most massive objects.  
Although a BS-GW-capture binary may be bound to the third object if the GW kick is small \citep{2019MNRAS.482...30S}, 
we ignore the third body after the binary formation for simplicity. 

\subsection{Evolution of binary separation, mass, etc.}
\label{sec:evolution_binary}

\subsubsection{Gaseous processes}

The velocity of BHs relative to the local motion of the AGN disk decreases due to the accretion torque and gas dynamical friction. 
The semi-major axis ($s$) evolves due to gas dynamical friction by the AGN disk and type I/II migration torques exerted by a compact circumbinary disk that forms within the Hill sphere of the binary. 
The radial distances of BHs from the SMBH ($r$) are also allowed to evolve due to type I/II torques from the AGN disk. 
Gas accretion affects BH masses, BH spins, and the orbital angular momentum directions of binaries (Paper~II). 
When gaps form around BHs, we assume that migration torques, gas dynamical friction, and gas accretion are weakened.  
Objects usually accumulate in gap forming regions (Paper~I), which act like migration traps. 

\subsubsection{Dynamical processes}
\label{sec:dynamical_effects}

We also account for dynamical interactions with single stars/BHs and BH binaries in a way similar to the Monte-Carlo method (Paper~I).  The binaries' semi-major axes, 
velocities, and orbital angular momentum directions evolve due to BS interactions, and the velocities of all BHs additionally evolve due to scattering. In particular, the binary separation decreases during hard BS interactions, and the single object and the binary receive a recoil kick. 
When the interactions are confined in 2D, kicks are also assigned within the AGN-disk plane. 

While binary component exchanges were neglected during BS interactions in Papers~I/II, 
here we assume that the components are always replaced by the most massive pair during hard-BS interactions, in which the binary becomes harder. 
During an exchange, the binary's binding energy is assumed to be unchanged
\citep{Sigurdsson93}. 
See the Appendix for the impact of including these exchange interactions.
In binary-binary interactions, we assume that the binary composed of the two most massive objects experiences two BS interactions with the two less massive objects, which become unbound after the interaction.

\subsubsection{Merger prescription}

Once binaries become sufficiently tight, their separation 
shrinks promptly by GW emission. 
We assume that the BHs merge when their pericenter distance becomes smaller than the innermost stable circular orbit, 
and then assign a kick velocity and mass loss due to the GW emission and prescribe BH spin evolution at the merger as in Paper~II.

\subsection{Binary eccentricity evolution}
\label{sec:ecc_ev}
After formation, the binary eccentricity $e$ changes due to GW emission, 
the torque by the circumbinary disk, and BS interactions. 
Below we describe our prescriptions for calculating those processes, 
which are newly incorporated in our model. 

For the GW emission, we follow \citet{Peters64} to track the change of the eccentricity.

For the torque by the circumbinary disk, 
we use the fitting function of Eq.~(2) of \citet{Zrake20} in $e\leq0.8$ 
and assume $de/d\mathrm{log}m_\mathrm{bin}=-2.3$ in $e>0.8$. 
Their results using 2D hydrodynamical simulations suggest disks drive binaries to $e\approx 0.45$ (see also \citealt{Munoz2019}).

During hard-BS interactions, we draw $e$ randomly from $f(e)$ (see below Eq.~\ref{eq:p_cap1n}). 
In the fiducial model, we assume that  
$e$ follows $f_{\rm 3D}(e)$ after all hard-BS interactions, but see different choices in the Appendix.

\begin{table*}
\begin{center}
	\caption{
    The assumptions and results in different models. 
    The input columns show the model number and the differences with respect to the fiducial model. 
    ``BS in 2$\rightarrow$3D'' represents the model 
    in which BS interactions occur in 2$\rightarrow$3D space ($\S\,\ref{sec:numerical_choices}$), 
    ``No BS'' represents the model in which BS interactions do not operate. 
    The output columns list the detection fraction of mergers among non GWC binaries ($f_\mathrm{noGWC}$), GWC binaries formed in SS encounters ($f_\mathrm{GWSS}$), and those in BS interactions ($f_\mathrm{GWBS}$) with respect to all mergers, 
    the detection fractions of mergers whose  $e_{10\,\mathrm{Hz}}$ are in the range $0$--$0.03$, $0.03$--$0.3$, $0.3$--$0.9$, and $0.9$--$1$, respectively, and the number of mergers ($N_\mathrm{mer}$). 
        }
\label{table_results}
\begin{tabular}{c|c||c|c|c|c|c|c|c|c}
\hline
\multicolumn{2}{c}{input} \vline& \multicolumn{8}{c}{output}\\\hline
Model&Parameter
&$f_\mathrm{noGWC}$
&$f_\mathrm{GWSS}$&$f_\mathrm{GWBS}$
&$f_\mathrm{e,0-003}$
&$f_\mathrm{e,003-03}$
&$f_\mathrm{e,03-09}$
&$f_\mathrm{e,09-1}$
&$N_\mathrm{mer}$
\\\hline

M1&Fiducial&
0.93&$0.02$&0.05
&0.86&0.065&0.022&0.057
&$1.1\times 10^3$
\\\hline

M2&BS in 2$\rightarrow$3D&
0.14&$3\times10^{-3}$&0.86
&0.25&0.082&0.20&0.47
&$2.3\times 10^3$
\\\hline

M3&No BS&
0.46&$0.54$&0
&0.46&0.092&0.26&0.19
&$9.0\times 10^2$
\\\hline

M4&$N_\mathrm{BH,ini}=6000$, $\beta_\mathrm{v}=1$&
0.55&$0.38$&0.078
&0.54&0.27&0.048&0.15
&$1.3\times 10^{2}$
\\\hline

\end{tabular}
\end{center}
\end{table*}

\begin{figure*}\begin{center}
\includegraphics[width=190mm]{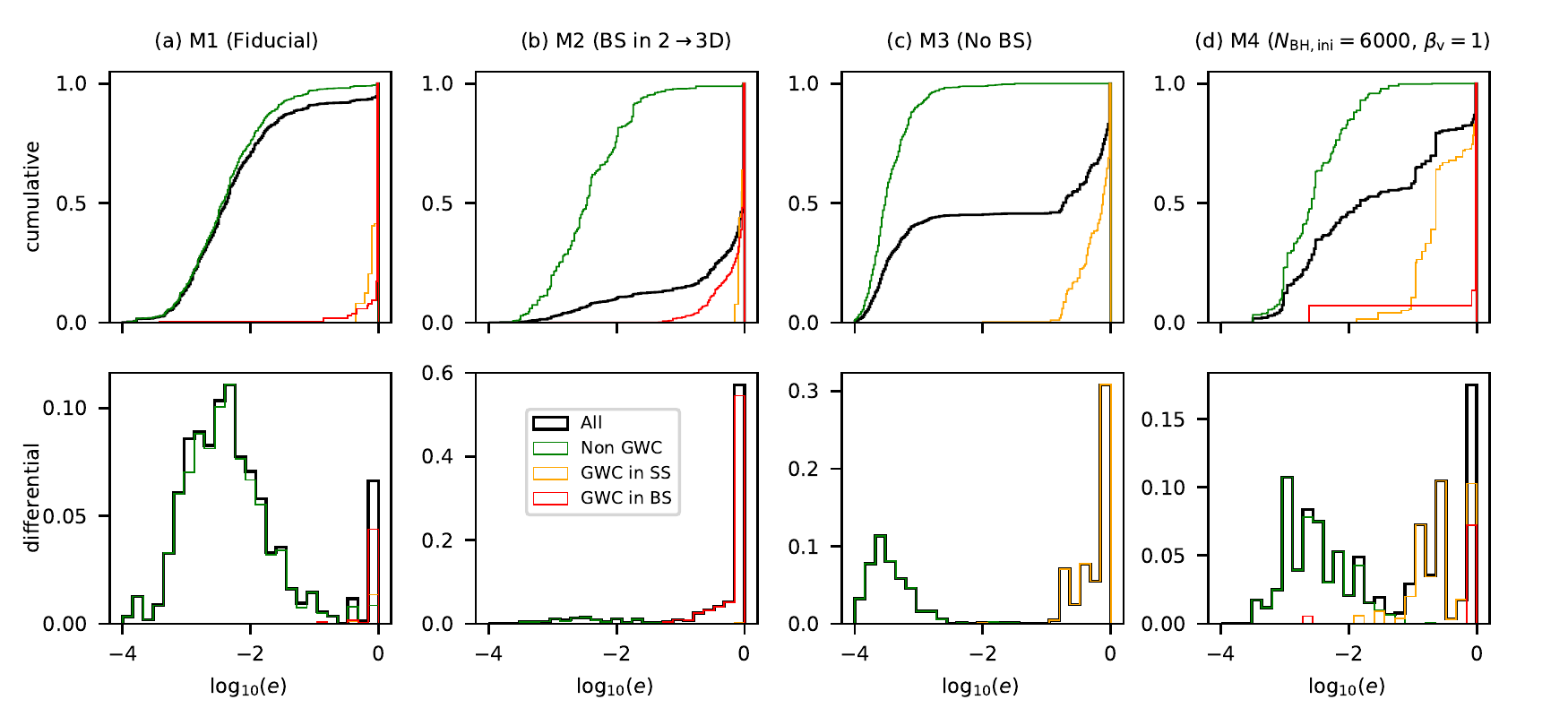}
\caption{
The cumulative (upper) and differential (lower) detection rate distributions of the eccentricity at 10Hz $e_\mathrm{10Hz}$ 
for models~M1 (a), M2 (b) in which BS interactions occur in 2$\rightarrow$3D space ($\S\,\ref{sec:numerical_choices}$), M3 (c) in which BS interaction does not operate, and M4 (d) in which the initial BH number is low ($N_\mathrm{ini}=6000$) and the initial velocity dispersion of BHs is high ($\beta_\mathrm{v}=1$). 
In addition to the sum of all binary mergers (black), separate contributions 
from binaries due to the non-GWC (green), SS-GWC (orange), and BS-GWC (red) are shown.
}\label{fig:model12}
\end{center}\end{figure*}

\begin{figure*}
\includegraphics[width=190mm]{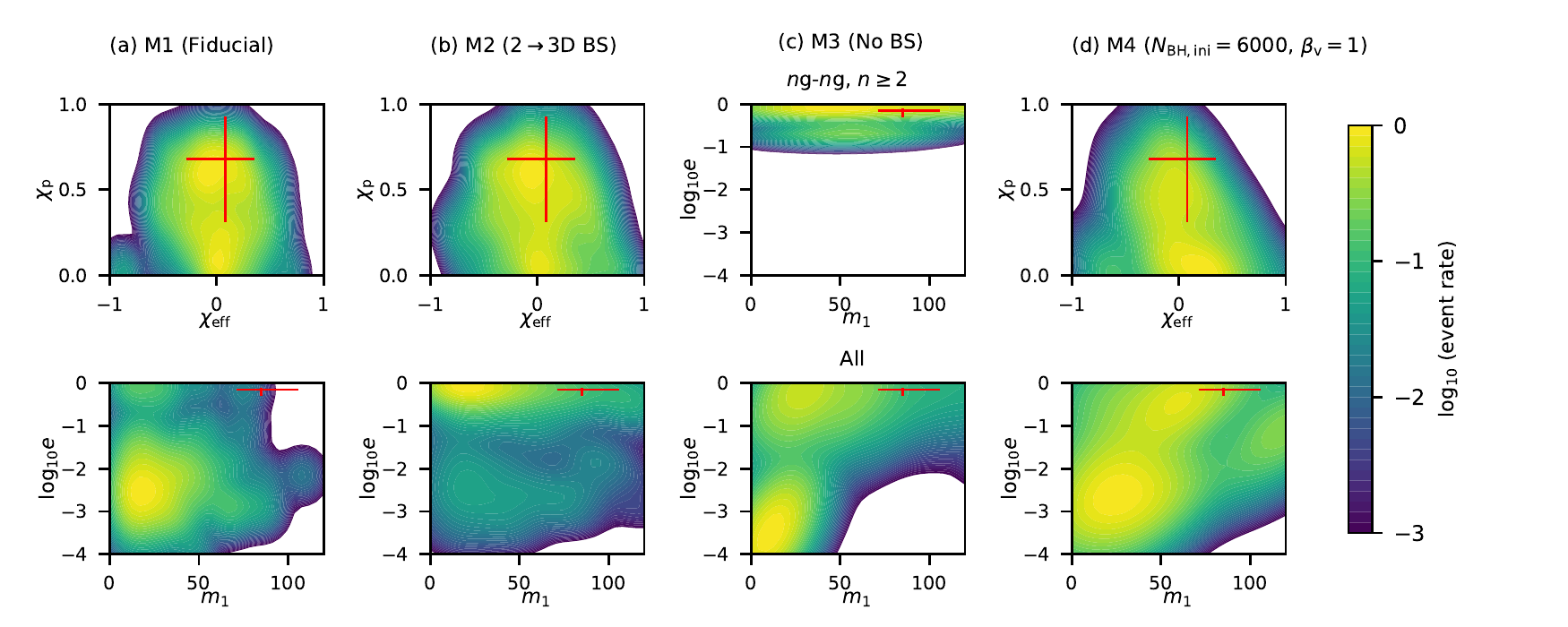}
\caption{Detection rate distributions of several parameters 
for the same models as in Fig.~\ref{fig:model12}. 
Upper and lower panels show distributions of $\chi_\mathrm{eff}$--$\chi_\mathrm{p}$ 
and $m_1$--$\mathrm{log}_{10} e_\mathrm{10Hz}$, respectively, while in model~M3, $m_1$--$\mathrm{log}_{10} e_\mathrm{10Hz}$ are shown for all mergers (lower) and mergers among $n$-generation ($n\geq2$) BHs (upper). 
The error bars correspond to the $90$ percentile credible intervals for GW190521 \citep{LIGO20_GW190521}, while we here roughly assume  $e_\mathrm{10Hz}=0.7^{+0.1}_{-0.2}$ referring to  \citet{Gayathri20_GW190521}. 
}\label{fig:spins_ecc}
\end{figure*}

\subsection{Models examined}

\label{sec:numerical_choices}

The fiducial model parameters (M1) are listed in Table~\ref{table:parameter_model}, which are the same as those in Papers~I/II. 
We also study the additional models~M2--M4, which differ in the treatment of the BS interactions.  
While 3D BS interactions are assumed in model~M1,  
motions of the three bodies may be confined to the AGN-disk plane before the interaction. 
Even in the latter case, the motion may be isotropized in 3D 
during the interaction by small perturbations, such as inhomogeneities/warps of the AGN disk and 
presence of other objects.
To mimic such 2$\rightarrow$3D interactions, we examine another model (M2) in which 
the capture probability during a BS interaction is given by Eq.~\eqref{eq:p_cap_disk} assuming that 
$N_\mathrm{int}=20$ 2D interactions take place with $f_{\rm 2D}(e)$. If a GW capture does not take place, the $e$ distribution is drawn from $f_{\rm 3D}(e)$ and the binary angular momentum direction is randomized due to 3D interactions in later phases. 
We neglect the chance for GWC after interactions are randomized to 3D as its probability is much lower than that during 2D interactions. 
The reason to adopt the model with 2$\rightarrow$3D BS interactions instead of 2D BS interactions is stated in Appendix. 
To assess the importance of BS interactions, we omit it altogether in model~M3.
Similar to model~M3, but more realistic is model~M4, 
in which BS interactions are made inefficient 
by reducing the initial BH number ($N_\mathrm{BH,ini}=6\times 10^3$) and enhancing the initial BH velocity dispersion ($\beta_\mathrm{v}=1$). 

In models~M1--M3, the initial number of 
BHs (binaries) is $2\times 10^4$ ($1.5\times 10^3$), while in model~M4 they are $6\times 10^3$ ($4.5\times 10^2$). We present the results at $10\,\mathrm{Myr}$.

\section{Results}
\label{sec:dependence}

In Fig.~\ref{fig:model12}, we show the eccentricity distribution of mergers and the contribution of individual binary formation channels to it.  
Fig.~\ref{fig:spins_ecc} shows the expected distribution of the mass-weighted sum of the individual spin components perpendicular to the orbital plane $\chi_{\rm eff}$, 
the precession parameter $\chi_{\rm p}$ (upper panels), the primary mass $m_{1}$, and  $e$ (lower panels) of 
the observed mergers except for model~M3.  For M3, the $\chi_\mathrm{eff}$ and $\chi_\mathrm{p}$ distributions are not shown since $\chi_\mathrm{p}$ is always 0 due to the assumption that the orbital planes of SS-GWC binaries are aligned with the 
AGN disk plane  ($\S\,\ref{sec:GW_SS}$). 
The merger rate is weighted by the detection volume in the same manner as in Papers~I/II and the eccentricity is calculated at the GW frequency of $10\,\mathrm{Hz}$ as in Eq.~(36) of \citet{Wen03}. 

In the fiducial model~M1, where the BS interaction is assumed to be 3D, 
$\sim14\%$ of mergers have a significantly nonzero eccentricity ($e_\mathrm{10Hz} \gtrsim 0.03$) measurable by the current ground-based detectors LIGO/Virgo/KAGRA (e.g. \citealt{Romero-Shaw19}, but see also \citealt{Huerta13}), 
due to GWC binary formation (Fig.~\ref{fig:model12} and Table~\ref{table_results}) and frequent BS interactions. 
While most AGN merger events take place among binaries of non-GWC origin, which have lower $e$ (Fig.~\ref{fig:model12} a), LISA observations will be able to measure $e$ as low as $e\gtrsim 10^{-3}-10^{-2}$ at $\sim 0.01\,\mathrm{Hz}$
\citep[e.g.][]{Nishizawa16}, corresponding to $e_\mathrm{10Hz}\gtrsim 10^{-7}$ if binaries evolve only by GW emission. 
$e_\mathrm{10Hz}$ are typically distributed in higher values ($\sim 10^{-3}$--$10^{-2}$) than those of mergers in globular clusters ($\sim 10^{-7}$--$10^{-4}$; \citealt{Rodriguez18bR}) and isolated binaries ($\sim 10^{-7}$--$10^{-6}$; \citealt{Breivik16}), while the similar values to those in triple system ($\sim 10^{-3}$--$10^{-2}$; \citealt{Antonini17}).

Due to randomization of the orbital angular momentum direction by frequent BS interactions before mergers, $\chi_\mathrm{eff}$, and $\chi_\mathrm{p}$ tend to be distributed at low and high values (upper panel, Fig.~\ref{fig:spins_ecc} a), respectively. 
The physical properties, $\chi_\mathrm{eff}$, $\chi_\mathrm{p}$, and $m_1$ of GW190521 
are generally consistent with the predictions by model~M1, 
while the possible high $e$ implied by \citet{Gayathri20_GW190521} is 
in tension with this model (lower panel, Fig.~\ref{fig:spins_ecc} a).

In model~M2 with 
2$\rightarrow$3D BS interactions, 
the fraction of the BS-GWC binary mergers among the detectable events  $f_\mathrm{GWBS}$ is elevated to 0.84 (from 0.10 of model~M1; see Table~\ref{table_results}), and they have very high eccentricity $e \gtrsim 0.1$ (Fig.~\ref{fig:model12} b). In this way, BS interactions significantly
increase 
the rate of highly eccentric mergers (see also \citealt{Samsing20}). 
The mass of GW190521, however, appears to be somewhat too high compared with the typical values in model~M2 (lower panel~b of Fig.~\ref{fig:spins_ecc}), but the model may still be consistent with the event as the mass distribution for high-$e$ events has been poorly constrained.

In model~M3, where the BS interaction is omitted,  
the (SS-)GWC binary mergers occupy a high fraction of $f_\mathrm{GWSS}=0.54$ favoring mergers between high-generation BHs (upper panel~c of Fig.~\ref{fig:spins_ecc}). 
High-generation BHs are those that have already experienced mergers in the past, 
and they tend to have high eccentricities around $\gtrsim 0.1$ (Fig.~\ref{fig:model12} c). 
Here, binary formation occurs mostly by the (SS-)GWC
because BHs rapidly migrate without delay by kicks at the BS interactions to the inner AGN disk of $r\lesssim 10^{-3}\,\mathrm{pc}$ (e.g. Fig.~1 of Paper~II), where non-GW-capture channels by the gas-capture and dynamical binary formation are inefficient due to high shear velocity ($\gtrsim$ tens km/s). 

In general, the frequency of the SS-GWC $f_\mathrm{GWSS}$ increases as the initial BH number decreases (see Appendix). 
For example, in model~M4, where $N_\mathrm{ini}=6000$ and $\beta_\mathrm{v}=1$, $f_\mathrm{GWSS}$ becomes as high as 0.38.
The prediction of 
positive $\chi_\mathrm{p}$, high $e$, and high-BH mass $m_1$ in this model agrees with the observed properties of GW190521 (Fig.~\ref{fig:spins_ecc}d). 
Although the number of mergers is smaller by a factor of $\sim10$ compared to that in the fiducial model, it is still consistent with the rate of GW190521-like events (see $\S\,5.5$ of Paper~I). 

As seen above, the AGN origin can naturally reproduce the occurrence of 
GW190521-like mergers. 
In several models, the eccentricity distribution allows us to distinguish specific types of the GWC events (orange and red lines in Fig.\ref{fig:model12}) due to differences in the initial relative velocity involved (Eq.~\ref{eq:rp_max}) 
and such high eccentricities are measurable by LIGO/Virgo/KAGRA. 
The consistency of other observables of GW190521 will be examined in a follow-up paper.

In the AGN channel, the SMBH helps retain recoiling merger remnants and allows multiple mergers among high-generation BHs, 
while the AGN disk helps to
(1) deliver the single objects to the dense inner nucleus where the interactions are frequent, (2) form binaries through the process of gas-capture binary formation, and (3) contributes to reducing the binary separation at relatively large separations. 
The rate of GWC mergers and the eccentricity distribution are influenced by the frequency of interactions in addition to the prescription for BS interactions, while they are less affected by other processes as discussed in more detail in the Appendix.

\section{Conclusions}

In this {\em Letter} we have investigated the eccentricity distribution of merging black holes (BHs) in active galactic nuclei (AGN) accretion disks. 
Our conclusions are summarized as follows:

\begin{enumerate}

\item
Due to frequent binary-single interactions in gap-forming regions of the AGN disk, eccentricities at $10\,\mathrm{Hz}$ 
are above $e_\mathrm{10Hz}\geq 10^{-4}$ for all stellar-mass BH mergers in AGNs in our models. 
LIGO together with LISA will be able to constrain the corresponding astrophysical models at high and low frequencies, respectively. 

\item 
If binary-single interactions occur in isotropic-three dimensional (3D) directions, 
$\sim8$--$30\%$ of detectable mergers have $e_\mathrm{10Hz}\gtrsim 0.03$, i.e. above the value possibly measurable by LIGO/Virgo/KAGRA.

\item 
If binary-single interactions occur in 2D or 2$\rightarrow$3D directions due to torques from the AGN disk, 
the mergers among gravitational wave capture binaries formed during BS interactions become very frequent ($\sim 15$--$90\%$). 
Also, if BHs can migrate to $r\lesssim10^{-3}\,\mathrm{pc}$ due to low BH density in the AGN disk, those formed in single-single encounters are significantly enhanced ($\sim 40$--$70\%$). 
In these cases, 
$\sim 10$--$70\%$ of detectable mergers have $e_\mathrm{10Hz}\geq 0.3$, which may explain the possible extreme eccentricity implied for GW190521 by \citet{Gayathri20_GW190521}.

\end{enumerate}

We have neglected the evolution of $e$ by soft-BS interactions, weak scattering, and gas dynamical friction, which presumably have minor effects on $e$ as their contributions on the later evolution of binaries are negligibly small. However, hierarchical triples composed of three stellar-mass compact objects \citep[e.g.][]{Antonini17}, 
neglected in this study, may significantly affect the $e$ distribution. 
We also have neglected other processes including 
the formation and evolution of compact objects (and their binaries) other than BHs, and the possible presence of massive perturbers \citep{Deme20}, which may affect various properties of mergers. 
Also, BS interactions for nearly 2D cases need to be elucidated using $N$-body simulations. 
These effects will be investigated in future work (see also Table~\ref{table:assumptions} in the Appendix).

\acknowledgments

This work is financially supported by the Grants-in-Aid for Basic Research by the Ministry of Education, Science and Culture of Japan (HT:17H01102, 17H06360, KO:17H02869, 17H01102, 17H06360).
ZH acknowledges support from NASA grant NNX15AB19G and NSF grants AST-1715661 and AST-2006176.
Simulations and analyses were carried out on Cray XC50 and computers at the Center for Computational Astrophysics, National Astronomical Observatory of Japan. 
This project has received funding from the European Research Council (ERC) under the European Union's Horizon 2020 research and innovation programme ERC-2014-STG under grant agreement No 638435 (GalNUC) to BK.

\bibliographystyle{yahapj.bst}
\bibliography{agn_bhm}

\begin{thebibliography}{}
\providecommand\natexlab[1]{#1}
\providecommand\JournalTitle[1]{#1}

\bibitem[{{Abbott} {et~al.}(2019{\natexlab{a}}){Abbott}, {et~al.}, {LIGO
  Scientific Collaboration}, \& {Virgo Collaboration}}]{TheLIGO18}
{Abbott}, B.~P., {et~al.}, {LIGO Scientific Collaboration}, \& {Virgo
  Collaboration}. 2019{\natexlab{a}},
  \href{http://dx.doi.org/10.1103/PhysRevX.9.031040}{\JournalTitle{Physical
  Review X}, 9, 031040}

\bibitem[{{Abbott} {et~al.}(2019{\natexlab{b}}){Abbott}, {Abbott}, {Abbott},
  {Abraham}, {Acernese}, {Ackley}, {Adams}, {Adhikari}, {Adya}, {Affeldt},
  {Agathos}, {Agatsuma}, {Aggarwal}, \& {Aguiar}}]{Abbott19_Ecc}
{Abbott}, B.~P., {Abbott}, R., {Abbott}, T.~D., {et~al.} 2019{\natexlab{b}},
  \href{http://dx.doi.org/10.3847/1538-4357/ab3c2d}{\JournalTitle{\apj}, 883,
  149}

\bibitem[{{Abbott} {et~al.}(2020{\natexlab{a}}){Abbott}, {Abbott}, {Abraham},
  {Acernese}, {Ackley}, {Adams}, {Adhikari}, {et~al.}}]{LIGO20_GW190521}
{Abbott}, R., {Abbott}, T.~D., {Abraham}, S., {et~al.} 2020{\natexlab{a}},
  \JournalTitle{arXiv e-prints}, arXiv:2009.01075

\bibitem[{{Abbott} {et~al.}(2020{\natexlab{b}}){Abbott}, {Abbott}, {Abraham},
  {Acernese}, {Ackley}, {Adams}, {Adhikari}, {et~al.}}]{LIGO20_GW190521_astro}
---. 2020{\natexlab{b}},
  \href{http://dx.doi.org/10.3847/2041-8213/aba493}{\JournalTitle{\apjl}, 900,
  L13}

\bibitem[{{Abbott} {et~al.}(2020{\natexlab{c}}){Abbott}, {Abbott}, {Abraham},
  {Acernese}, {Ackley}, {Adams}, {et~al.}}]{LIGO20_GW190412}
---. 2020{\natexlab{c}}, \JournalTitle{arXiv e-prints}, arXiv:2004.08342

\bibitem[{{Abbott} {et~al.}(2020{\natexlab{d}}){Abbott}, {Abbott}, {Abraham},
  {Acernese}, {Ackley}, {Adams}, {et~al.}}]{LIGO20_GW190814}
---. 2020{\natexlab{d}},
  \href{http://dx.doi.org/10.3847/2041-8213/ab960f}{\JournalTitle{\apjl}, 896,
  L44}

\bibitem[{{Antonini} \& {Perets}(2012)}]{Antonini12}
{Antonini}, F., \& {Perets}, H.~B. 2012,
  \href{http://dx.doi.org/10.1088/0004-637X/757/1/27}{\JournalTitle{\apj}, 757,
  27}

\bibitem[{{Antonini} {et~al.}(2017){Antonini}, {Toonen}, \&
  {Hamers}}]{Antonini17}
{Antonini}, F., {Toonen}, S., \& {Hamers}, A.~S. 2017,
  \href{http://dx.doi.org/10.3847/1538-4357/aa6f5e}{\JournalTitle{\apj}, 841,
  77}

\bibitem[{{Arca-Sedda} {et~al.}(2018){Arca-Sedda}, {Li}, \&
  {Kocsis}}]{ArcaSedda18}
{Arca-Sedda}, M., {Li}, G., \& {Kocsis}, B. 2018, \JournalTitle{arXiv
  e-prints}, arXiv:1805.06458

\bibitem[{{Bartos} {et~al.}(2017){Bartos}, {Kocsis}, {Haiman}, \&
  {M{\'a}rka}}]{Bartos17}
{Bartos}, I., {Kocsis}, B., {Haiman}, Z., \& {M{\'a}rka}, S. 2017,
  \href{http://adsabs.harvard.edu/abs/2017ApJ...835..165B}{\JournalTitle{\apj},
  835, 165}

\bibitem[{{Breivik} {et~al.}(2016){Breivik}, {Rodriguez}, {Larson}, {Kalogera},
  \& {Rasio}}]{Breivik16}
{Breivik}, K., {Rodriguez}, C.~L., {Larson}, S.~L., {Kalogera}, V., \& {Rasio},
  F.~A. 2016,
  \href{http://dx.doi.org/10.3847/2041-8205/830/1/L18}{\JournalTitle{\apjl},
  830, L18}

\bibitem[{{Deme} {et~al.}(2020){Deme}, {Meiron}, \& {Kocsis}}]{Deme20}
{Deme}, B., {Meiron}, Y., \& {Kocsis}, B. 2020,
  \href{http://dx.doi.org/10.3847/1538-4357/ab7921}{\JournalTitle{\apj}, 892,
  130}

\bibitem[{{Farmer} {et~al.}(2019){Farmer}, {Renzo}, {de Mink}, {Marchant}, \&
  {Justham}}]{Farmer19}
{Farmer}, R., {Renzo}, M., {de Mink}, S.~E., {Marchant}, P., \& {Justham}, S.
  2019, \href{http://dx.doi.org/10.3847/1538-4357/ab518b}{\JournalTitle{\apj},
  887, 53}

\bibitem[{{Fragione} \& {Bromberg}(2019)}]{Fragione_Bromberg2019}
{Fragione}, G., \& {Bromberg}, O. 2019,
  \href{http://dx.doi.org/10.1093/mnras/stz2024}{\JournalTitle{\mnras}, 488,
  4370}

\bibitem[{{Fragione} {et~al.}(2019){Fragione}, {Grishin}, {Leigh}, {Perets}, \&
  {Perna}}]{Fragione+2019}
{Fragione}, G., {Grishin}, E., {Leigh}, N. W.~C., {Perets}, H.~B., \& {Perna},
  R. 2019,
  \href{http://dx.doi.org/10.1093/mnras/stz1651}{\JournalTitle{\mnras}, 488,
  47}

\bibitem[{{Fuller} \& {Ma}(2019)}]{Fuller19_massive}
{Fuller}, J., \& {Ma}, L. 2019,
  \href{http://dx.doi.org/10.3847/2041-8213/ab339b}{\JournalTitle{\apjl}, 881,
  L1}

\bibitem[{{Gayathri} {et~al.}(2020){Gayathri}, {Healy}, {Lange}, {O'Brien},
  {Szczepanczyk}, {Bartos}, {Campanelli}, {Klimenko}, {Lousto}, \&
  {O'Shaughnessy}}]{Gayathri20_GW190521}
{Gayathri}, V., {Healy}, J., {Lange}, J., {et~al.} 2020, \JournalTitle{arXiv
  e-prints}, arXiv:2009.05461

\bibitem[{{Gond{\'a}n} \& {Kocsis}(2019)}]{Gondan_Kocsis19}
{Gond{\'a}n}, L., \& {Kocsis}, B. 2019,
  \href{http://dx.doi.org/10.3847/1538-4357/aaf893}{\JournalTitle{\apj}, 871,
  178}

\bibitem[{{Gond{\'a}n} \& {Kocsis}(2020)}]{Gondan20_GW190521}
---. 2020, \JournalTitle{arXiv e-prints}, arXiv:2011.02507

\bibitem[{{Gond{\'a}n} {et~al.}(2018){Gond{\'a}n}, {Kocsis}, {Raffai}, \&
  {Frei}}]{Gondan18a}
{Gond{\'a}n}, L., {Kocsis}, B., {Raffai}, P., \& {Frei}, Z. 2018,
  \href{http://dx.doi.org/10.3847/1538-4357/aabfee}{\JournalTitle{\apj}, 860,
  5}

\bibitem[{{Graham} {et~al.}(2020){Graham}, {Ford}, {McKernan}, {Ross}, {Stern},
  {Burdge}, {Coughlin}, {Djorgovski}, {Drake}, {Duev}, {Kasliwal}, {Mahabal},
  {van Velzen}, {Belecki}, {Bellm}, {Burruss}, {Cenko}, {Cunningham}, {Helou},
  {Kulkarni}, {Masci}, {Prince}, {Reiley}, {Rodriguez}, {Rusholme}, {Smith}, \&
  {Soumagnac}}]{Graham20}
{Graham}, M.~J., {Ford}, K.~E.~S., {McKernan}, B., {et~al.} 2020,
  \href{http://dx.doi.org/10.1103/PhysRevLett.124.251102}{\JournalTitle{\prl},
  124, 251102}

\bibitem[{{Heggie}(1975)}]{Heggie75}
{Heggie}, D.~C. 1975,
  \href{http://adsabs.harvard.edu/abs/1975MNRAS.173..729H}{\JournalTitle{\mnras},
  173, 729}

\bibitem[{{Huerta} \& {Brown}(2013)}]{Huerta13}
{Huerta}, E.~A., \& {Brown}, D.~A. 2013,
  \href{http://dx.doi.org/10.1103/PhysRevD.87.127501}{\JournalTitle{Physical
  Review D}, 87, 127501}

\bibitem[{{Kocsis} {et~al.}(2006){Kocsis}, {G{\'a}sp{\'a}r}, \&
  {M{\'a}rka}}]{2006ApJ...648..411K}
{Kocsis}, B., {G{\'a}sp{\'a}r}, M.~E., \& {M{\'a}rka}, S. 2006,
  \href{http://dx.doi.org/10.1086/505641}{\JournalTitle{\apj}, 648, 411}

\bibitem[{{Leigh} {et~al.}(2018){Leigh}, {Geller}, {McKernan}, {Ford}, {Mac
  Low}, {Bellovary}, {Haiman}, {Lyra}, {Samsing}, {O'Dowd}, {Kocsis}, \&
  {Endlich}}]{Leigh18}
{Leigh}, N.~W.~C., {Geller}, A.~M., {McKernan}, B., {et~al.} 2018,
  \href{http://dx.doi.org/10.1093/mnras/stx3134}{\JournalTitle{\mnras}, 474,
  5672}

\bibitem[{{Lower} {et~al.}(2018){Lower}, {Thrane}, {Lasky}, \&
  {Smith}}]{Lower18}
{Lower}, M.~E., {Thrane}, E., {Lasky}, P.~D., \& {Smith}, R. 2018,
  \href{http://dx.doi.org/10.1103/PhysRevD.98.083028}{\JournalTitle{\prd}, 98,
  083028}

\bibitem[{{Lubow} {et~al.}(1999){Lubow}, {Seibert}, \& {Artymowicz}}]{Lubow99}
{Lubow}, S.~H., {Seibert}, M., \& {Artymowicz}, P. 1999,
  \href{http://adsabs.harvard.edu/abs/1999ApJ...526.1001L}{\JournalTitle{\apj},
  526, 1001}

\bibitem[{{Martinez} {et~al.}(2020){Martinez}, {Fragione}, {Kremer},
  {Chatterjee}, {Rodriguez}, {Samsing}, {Ye}, {Weatherford}, {Zevin}, {Naoz},
  \& {Rasio}}]{Martinez+20}
{Martinez}, M. A.~S., {Fragione}, G., {Kremer}, K., {et~al.} 2020,
  \href{http://dx.doi.org/10.3847/1538-4357/abba25}{\JournalTitle{\apj}, 903,
  67}

\bibitem[{{McKernan} {et~al.}(2018){McKernan}, {Ford}, {Bellovary}, {Leigh},
  {Haiman}, {Kocsis}, {Lyra}, {Mac Low}, {Metzger}, {O'Dowd}, {Endlich}, \&
  {Rosen}}]{McKernan17}
{McKernan}, B., {Ford}, K.~E.~S., {Bellovary}, J., {et~al.} 2018,
  \href{http://adsabs.harvard.edu/abs/2018ApJ...866...66M}{\JournalTitle{\apj},
  866, 66}

\bibitem[{{McKernan} {et~al.}(2019){McKernan}, {Ford}, {Bartos}, {Graham},
  {Lyra}, {Marka}, {Marka}, {Ross}, {Stern}, \& {Yang}}]{2019ApJ...884L..50M}
{McKernan}, B., {Ford}, K.~E.~S., {Bartos}, I., {et~al.} 2019,
  \href{http://dx.doi.org/10.3847/2041-8213/ab4886}{\JournalTitle{\apjl}, 884,
  L50}

\bibitem[{{Mu{\~n}oz} {et~al.}(2019){Mu{\~n}oz}, {Miranda}, \&
  {Lai}}]{Munoz2019}
{Mu{\~n}oz}, D.~J., {Miranda}, R., \& {Lai}, D. 2019,
  \href{http://dx.doi.org/10.3847/1538-4357/aaf867}{\JournalTitle{\apj}, 871,
  84}

\bibitem[{{Nishizawa} {et~al.}(2016){Nishizawa}, {Berti}, {Klein}, \&
  {Sesana}}]{Nishizawa16}
{Nishizawa}, A., {Berti}, E., {Klein}, A., \& {Sesana}, A. 2016,
  \href{http://adsabs.harvard.edu/abs/2016PhRvD..94f4020N}{\JournalTitle{Phys.
  Rev. D}, 94, 064020}

\bibitem[{{O'Leary} {et~al.}(2009){O'Leary}, {Kocsis}, \& {Loeb}}]{OLeary09}
{O'Leary}, R.~M., {Kocsis}, B., \& {Loeb}, A. 2009,
  \href{http://adsabs.harvard.edu/abs/2009MNRAS.395.2127O}{\JournalTitle{\mnras},
  395, 2127}

\bibitem[{{Peters}(1964)}]{Peters64}
{Peters}, P.~C. 1964, \JournalTitle{Phys. Rev.}, 136, 1224

\bibitem[{{Rasskazov} \& {Kocsis}(2019)}]{Rasskazov19}
{Rasskazov}, A., \& {Kocsis}, B. 2019,
  \href{http://dx.doi.org/10.3847/1538-4357/ab2c74}{\JournalTitle{\apj}, 881,
  20}

\bibitem[{{Rodriguez} {et~al.}(2018){Rodriguez}, {Amaro-Seoane}, {Chatterjee},
  {Kremer}, {Rasio}, {Samsing}, {Ye}, \& {Zevin}}]{Rodriguez18bR}
{Rodriguez}, C.~L., {Amaro-Seoane}, P., {Chatterjee}, S., {et~al.} 2018,
  \href{http://dx.doi.org/10.1103/PhysRevD.98.123005}{\JournalTitle{\prd}, 98,
  123005}

\bibitem[{{Romero-Shaw} {et~al.}(2019){Romero-Shaw}, {Lasky}, \&
  {Thrane}}]{Romero-Shaw19}
{Romero-Shaw}, I.~M., {Lasky}, P.~D., \& {Thrane}, E. 2019, \JournalTitle{arXiv
  e-prints}, arXiv:1909.05466

\bibitem[{{Romero-Shaw} {et~al.}(2020){Romero-Shaw}, {Lasky}, {Thrane}, \&
  {Calderon Bustillo}}]{Romero-Shaw20}
{Romero-Shaw}, I.~M., {Lasky}, P.~D., {Thrane}, E., \& {Calderon Bustillo}, J.
  2020, \JournalTitle{arXiv e-prints}, arXiv:2009.04771

\bibitem[{{Samsing}(2018)}]{Samsing18b}
{Samsing}, J. 2018,
  \href{http://dx.doi.org/10.1103/PhysRevD.97.103014}{\JournalTitle{\prd}, 97,
  103014}

\bibitem[{{Samsing} \& {D'Orazio}(2018)}]{2018MNRAS.481.5445S}
{Samsing}, J., \& {D'Orazio}, D.~J. 2018,
  \href{http://dx.doi.org/10.1093/mnras/sty2334}{\JournalTitle{\mnras}, 481,
  5445}

\bibitem[{{Samsing} {et~al.}(2020{\natexlab{a}}){Samsing}, {D'Orazio},
  {Kremer}, {Rodriguez}, \& {Askar}}]{2020PhRvD.101l3010S}
{Samsing}, J., {D'Orazio}, D.~J., {Kremer}, K., {Rodriguez}, C.~L., \& {Askar},
  A. 2020{\natexlab{a}},
  \href{http://dx.doi.org/10.1103/PhysRevD.101.123010}{\JournalTitle{\prd},
  101, 123010}

\bibitem[{{Samsing} \& {Ilan}(2019)}]{2019MNRAS.482...30S}
{Samsing}, J., \& {Ilan}, T. 2019,
  \href{http://dx.doi.org/10.1093/mnras/sty2249}{\JournalTitle{\mnras}, 482,
  30}

\bibitem[{{Samsing} {et~al.}(2014){Samsing}, {MacLeod}, \&
  {Ramirez-Ruiz}}]{Samsing14}
{Samsing}, J., {MacLeod}, M., \& {Ramirez-Ruiz}, E. 2014,
  \href{http://dx.doi.org/10.1088/0004-637X/784/1/71}{\JournalTitle{\apj}, 784,
  71}

\bibitem[{{Samsing} {et~al.}(2020{\natexlab{b}}){Samsing}, {Bartos},
  {D'Orazio}, {Haiman}, {Kocsis}, {Leigh}, {Liu}, {Pessah}, \&
  {Tagawa}}]{Samsing20}
{Samsing}, J., {Bartos}, I., {D'Orazio}, D.~J., {et~al.} 2020{\natexlab{b}},
  \JournalTitle{arXiv e-prints}, arXiv:2010.09765

\bibitem[{{Sigurdsson} \& {Phinney}(1993)}]{Sigurdsson93}
{Sigurdsson}, S., \& {Phinney}, E.~S. 1993,
  \href{http://dx.doi.org/10.1086/173190}{\JournalTitle{\apj}, 415, 631}

\bibitem[{{Silsbee} \& {Tremaine}(2017)}]{Silsbee17}
{Silsbee}, K., \& {Tremaine}, S. 2017,
  \href{http://adsabs.harvard.edu/abs/2017ApJ...836...39S}{\JournalTitle{\apj},
  836, 39}

\bibitem[{{Stone} {et~al.}(2017){Stone}, {Metzger}, \& {Haiman}}]{Stone17}
{Stone}, N.~C., {Metzger}, B.~D., \& {Haiman}, Z. 2017,
  \href{http://dx.doi.org/10.1093/mnras/stw2260}{\JournalTitle{\mnras}, 464,
  946}

\bibitem[{{Szolgyen} \& {Kocsis}(2018)}]{Szolgyen18}
{Szolgyen}, A., \& {Kocsis}, B. 2018,
  \href{http://adsabs.harvard.edu/abs/2018PhRvL.121j1101S}{\JournalTitle{Phys.
  Rev. Lett.}, 121, 101101}

\bibitem[{{Tagawa} {et~al.}(2020{\natexlab{a}}){Tagawa}, {Haiman}, {Bartos}, \&
  {Kocsis}}]{Tagawa20b_spin}
{Tagawa}, H., {Haiman}, Z., {Bartos}, I., \& {Kocsis}, B. 2020{\natexlab{a}},
  \href{http://dx.doi.org/10.3847/1538-4357/aba2cc}{\JournalTitle{\apj}, 899,
  26}

\bibitem[{{Tagawa} {et~al.}(2020{\natexlab{b}}){Tagawa}, {Haiman}, \&
  {Kocsis}}]{Tagawa19}
{Tagawa}, H., {Haiman}, Z., \& {Kocsis}, B. 2020{\natexlab{b}},
  \href{http://dx.doi.org/10.3847/1538-4357/ab9b8c}{\JournalTitle{\apj}, 898,
  25}

\bibitem[{{Thompson} {et~al.}(2005){Thompson}, {Quataert}, \&
  {Murray}}]{Thompson05}
{Thompson}, T.~A., {Quataert}, E., \& {Murray}, N. 2005,
  \href{http://adsabs.harvard.edu/abs/2005ApJ...630..167T}{\JournalTitle{\apj},
  630, 167}

\bibitem[{{Valtonen} \& {Karttunen}(2006)}]{Valtonen06}
{Valtonen}, M., \& {Karttunen}, H. 2006,
  \href{http://adsabs.harvard.edu/abs/2006tbp..book.....V}{\JournalTitle{The
  Three-Body Problem (Cambridge: Cambridge University Press)}}

\bibitem[{{Wen}(2003)}]{Wen03}
{Wen}, L. 2003, \href{http://dx.doi.org/10.1086/378794}{\JournalTitle{\apj},
  598, 419}

\bibitem[{{Yang} {et~al.}(2020){Yang}, {Gayathri}, {Bartos}, {Haiman},
  {Safarzadeh}, \& {Tagawa}}]{Yang20_gap}
{Yang}, Y., {Gayathri}, V., {Bartos}, I., {et~al.} 2020, \JournalTitle{arXiv
  e-prints}, arXiv:2007.04781

\bibitem[{{Yang} {et~al.}(2019){Yang}, {Bartos}, {Gayathri}, {Ford}, {Haiman},
  {Klimenko}, {Kocsis}, {M{\'a}rka}, {M{\'a}rka}, {McKernan}, \&
  {O'Shaughnessy}}]{Yang19b_prl}
{Yang}, Y., {Bartos}, I., {Gayathri}, V., {et~al.} 2019,
  \href{http://dx.doi.org/10.1103/PhysRevLett.123.181101}{\JournalTitle{\prl},
  123, 181101}

\bibitem[{{Zackay} {et~al.}(2019){Zackay}, {Dai}, {Venumadhav}, {Roulet}, \&
  {Zaldarriaga}}]{Zackay19_GW170817A}
{Zackay}, B., {Dai}, L., {Venumadhav}, T., {Roulet}, J., \& {Zaldarriaga}, M.
  2019, \JournalTitle{arXiv e-prints}, arXiv:1910.09528

\bibitem[{{Zevin} {et~al.}(2019){Zevin}, {Samsing}, {Rodriguez}, {Haster}, \&
  {Ramirez-Ruiz}}]{Zevin19}
{Zevin}, M., {Samsing}, J., {Rodriguez}, C., {Haster}, C.-J., \&
  {Ramirez-Ruiz}, E. 2019,
  \href{http://adsabs.harvard.edu/abs/2019ApJ...871...91Z}{\JournalTitle{\apj},
  871, 91}

\bibitem[{{Zrake} {et~al.}(2020){Zrake}, {Tiede}, {MacFadyen}, \&
  {Haiman}}]{Zrake20}
{Zrake}, J., {Tiede}, C., {MacFadyen}, A., \& {Haiman}, Z. 2020,
  \JournalTitle{arXiv e-prints}, arXiv:2010.09707

\end{thebibliography}

\appendix

\section{Alternative models}
\label{sec:other_models}

\begin{figure*}\begin{center}
\includegraphics[width=180mm]{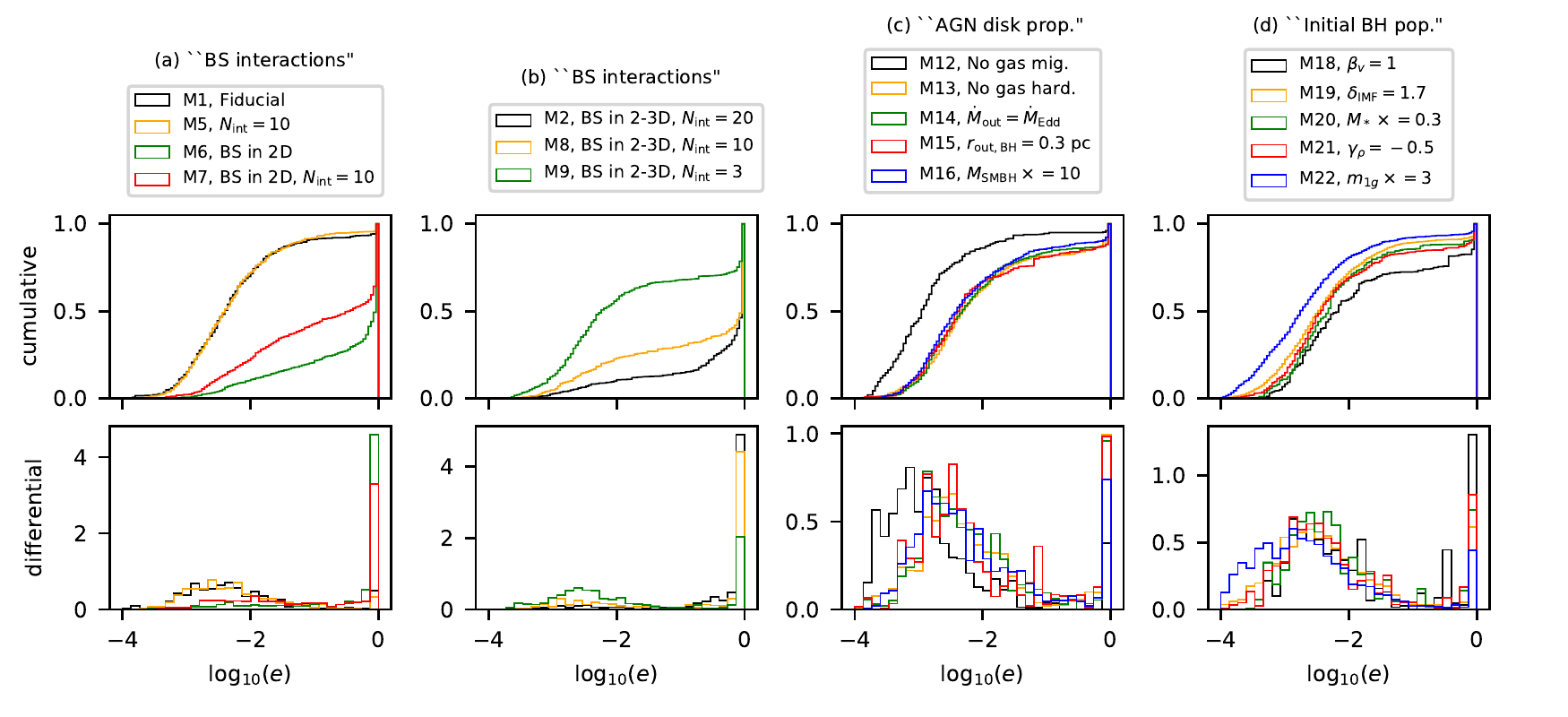}\caption{
The cumulative (upper) and differential (lower) detection rate distributions for the eccentricity at $10\,\mathrm{HZ}$ among all merging binaries for models~M1, M2, M5--M9, M12--M16, and M18--M22. Different colors and rows present the distributions for different variations of the models, as labeled on the upper legends and listed in Table~\ref{table_results}.}
\label{fig:models}\end{center}\end{figure*}

Here we present the dependence of the eccentricity distribution using a total of 26 different models, listed in Table~\ref{table_results_app}. 
Fig.~\ref{fig:models} shows the $e$ distributions for models~M1, M2, M5--M9, M12--M16, M18--M22. 
Panels~(a)--(b) show the dependence on $N_\mathrm{int}$ when BS interactions occur in isotropic-3D, 2D, or 2$\rightarrow$3D directions ($\S\,\ref{sec:numerical_choices}$).  
When BS interactions occur in isotropic-2D directions, we assume that $e$ follows $f_{\rm 2D}(e)$ 
after BS interactions if both the binary and the third body are in the AGN disk and $f_{\rm 3D}(e)$ otherwise (models~M6 and M7). 
The panels demonstrate that $N_\mathrm{int}$ significantly influences $f_\mathrm{GWBS}$ and the $e$ distribution of mergers (models~M1, M2, M5--M9). 
Here, the fiducial value of $N_\mathrm{int}\sim 20$ is motivated by numerical simulations in \citet{Samsing14} and \citet{Samsing20} for equal-mass and isotropic-3D or 2D cases. 

The eccentricity distributions for models with 2$\rightarrow$3D and 2D BS interactions are similar with each other (green lines in panels~a and black lines in panels~b), while $\chi_\mathrm{p}$ is always zero for 2D cases as we assume that binary orbital angular momentum directions are not randomized after 2D BS interactions. If this assumption is reasonable, models with pure 2D BS interactions may be difficult to produce GW190521 like merger, which has non-zero $\chi_\mathrm{p}$. 
As models with 2D BS interactions may be already excluded, we present the model with 2$\rightarrow$3D BS interactions in the main text, which is also possible to be occurring in AGN disks as discussed in $\S\,\ref{sec:numerical_choices}$.

We here discuss the feasibility of 2D (and 2D$\rightarrow$3D) interactions. 
If the velocities of BHs are damped due to interactions with gas before BS interactions, the interactions could be confined to a 2D plane. This is because the velocity of BHs perpendicular to the AGN disk $v_z$ is $\sim 0.01\,\mathrm{km/s}$ at $r \sim \mathrm{pc}$ after the velocity damping (see Fig.~5 in Paper~I), which is much smaller than $v_\mathrm{kep}\sim 100\,\mathrm{km/s}(r/\mathrm{pc})^{-1/2}(M_\mathrm{SMBH}/4\times 10^6\,\Msun)^{1/2}$. Referring to these values, the typical orbital elevation of BHs after damping is approximately $\sim  (v_z/v_\mathrm{kep})r \sim 10^{-4} r$ at $r\sim \mathrm{pc}$ and the Hill radius of BHs with $\sim 10\,\Msun$ is $\sim 10^{-2} r$. Thus, the inclination of orbits passing inside the Hill radius is $\sim 10^{-2}$, justifying the assumption that interactions might occur in a 2D plane, and our models with 2D or 2D$\rightarrow$3D interactions. 
On the other hand, around $r\sim 0.01\,\mathrm{pc}$, BS interactions occur soon after the binary is captured to the AGN disk (Fig.~5 in Paper~I), 
whose timescale is much shorter than the migration timescale ($t_\mathrm{mig}\gtrsim 0.3\,\mathrm{Myr}$, Fig.~10 of Paper~I). 
This implies that interacting objects orbit in similar annuli without migrating motion, and interactions occur while their velocities are still being damped. In this case, the directions of interactions likely become 3D. 
We conclude that the inclination angles during the BS interactions are roughly confined in a 2D plane in the outer regions of the AGN disk, while they approach a 3D distribution around $r\sim0.01\,\mathrm{pc}$, where the BH density is high. 
However, as also mentioned in the main text, possible inhomogeneities/warps of the AGN disk, ignored in our model, may weaken or diminish any alignment and suppress 2D (or 2D$\rightarrow$3D) interactions. As there are several uncertainties regarding this issue, further investigations would be required to understand this issue more thoroughly.

We also study several parameters 
related to the properties of the AGN disk in panels~(c) (models~M12--M17), the initial BH population in panels~(d) (models~M18--M24), and gaseous torque (model~M11). 
We find that 
the $e$ distribution is less affected by many parameters and processes, such as torque from circumbinary disks (model~M11), 
hardening due to gaseous processes (model~M13), 
the accretion rate from outer regions (model~M14), 
the size of AGN disks (model~M15), 
the mass of the SMBH (model~M16), 
the accretion rate onto stellar-mass BHs (model~M17), 
the slope of the initial mass function (model~M19), 
the total mass of stars (model~M20), 
or 
the slope of the initial radial distribution of BHs (model~M21). 
Indeed, the $e$ distribution is determined almost entirely by the rate of BS interactions and GWC binary formation. The important process regulating the rates is the migration of BHs to the dense inner regions where interactions are frequent.  

On the other hand, $e_{\rm 10 Hz}$ is confined to low values when radial migration does not operate (model~M12, black lines in panels~c of Fig.~\ref{fig:models}) or if the initial BH masses are $3\times$ larger (model~M22, blue lines in panels~d of Fig.~\ref{fig:models}). 
The frequency of BS interactions is influenced by these changes as follows. 
When radial migration does not operate (model~M12), 
BS interactions become less frequent as migration enhances the BH density in gap-forming inner regions 
where migration slows down and BHs accumulate (e.g. Paper~I). 
Also, in the outer regions, it takes longer for binaries kicked by BS interactions to be re-captured to the AGN disk. 
Then a longer time is available to the binaries to merge by GW emission, reducing the eccentricity at merger. 
Similar effects are expected 
when the initial BH masses are multiplied by 3 (model~M22). 
In this model, massive BHs open gaps in regions further out, 
and mergers occur at $r\sim 0.003-3\,\mathrm{pc}$ in model~M22, while those in the fiducial model are limited to $r\sim 0.003-0.03\,\mathrm{pc}$. 
As these parameters affect the eccentricity distribution at low $e$, these variations are expected to be constrained by LISA. 
Overall, capture of BHs into the AGN disk and migration to gap-forming inner regions increases the rate of BS interactions, which increases the typical values of $e_\mathrm{10Hz}$, in addition to prescriptions for BS interactions. On the other hand, the $e_\mathrm{10Hz}$ distribution is less affected by other processes.

Also, the detection fraction of mergers among SS-GWC binaries ($f_\mathrm{GWSS}$) increases as the BH density in the AGN disk is reduced (models~M4, M18, M23, M24). 
When the BH density in the AGN disk is low, BS interactions are infrequent, and then, BHs can easily migrate to the inner AGN disk where SS-GWC binary formation is efficient as discussed in the main text. 
It may be possible to constrain the efficiency of migration of BHs in the AGN disk by measuring $f_\mathrm{GWSS}$ by LIGO/Virgo/KAGRA. 

When the exchange at BS interactions is neglected (model~M26 or all models in Papers~I/II), light BHs easily reside in binaries and merge. 
As the energy required to be extracted from the binary to merge is smaller for lighter binaries, the number of mergers among non-GWC binaries is enhanced by ignoring the exchange (Table~\ref{table_results_app}).

\begin{table*}
\begin{center}
	\caption{
	Same with Table~\ref{table_results_app}, but results for all models. 
	  ``BS in 2D'' represents the model in which 
         BS interactions occur in 2D. 
     ``No BS'' represents the model in which BS interactions do not operate. 
     ``No gas torque on ecc'', 
     ``No gas hardening'', and ``No gas migration'' respectively represent the models in which the gaseous torques are neglected with respect to the evolution of $e$, 
     the binary semi-major axis, 
         and the radial position of the binary center of mass within the AGN disk. 
        ``$3\times m_{1g}$'' represents the model in which initial BH masses are multiplied by 3. 
``No exchange in BS'' represents the model in which the binary components are not exchanged during BS interactions. 
}
\label{table_results_app}
\begin{tabular}{c|c||c|c|c|c|c|c|c|c}
\hline
\multicolumn{2}{c}{input} \vline& \multicolumn{8}{c}{output}\\\hline
Model&Parameter
&$f_\mathrm{noGWC}$
&$f_\mathrm{GWSS}$&$f_\mathrm{GWBS}$
&$f_\mathrm{e,0-003}$
&$f_\mathrm{e,003-03}$
&$f_\mathrm{e,03-09}$
&$f_\mathrm{e,09-1}$
&$N_\mathrm{mer}$
\\\hline

M1&Fiducial&
0.93&$0.02$&0.05
&0.86&0.065&0.022&0.057
&$1.1\times 10^3$
\\\hline

M2&BS in 2$\rightarrow$3D&
0.14&$3\times10^{-3}$&0.86
&0.25&0.082&0.20&0.47
&$2.3\times 10^3$
\\\hline

M3&No BS&
0.46&$0.54$&0
&0.46&0.092&0.26&0.19
&$9.0\times 10^2$
\\\hline

M4&$N_\mathrm{BH,ini}=6000$, $\beta_\mathrm{v}=1$&
0.55&$0.38$&0.08
&0.54&0.27&0.048&0.15
&$1.3\times 10^{2}$
\\\hline

M5&$N_\mathrm{int}=10$&
0.96&$5\times 10^{-3}$&0.03
&0.85&0.10&0.0084&0.043
&$1.1\times 10^3$
\\\hline

M6&BS in 2D&
0.25&$0.02$&0.73
&0.25&0.10&0.19&0.46
&$3.1\times 10^3$
\\\hline

M7&BS in 2D, $N_\mathrm{int}=10$&
0.49&$0.01$&0.50
&0.41&0.13&0.13&0.34
&$2.8\times 10^3$
\\\hline

M8&BS in 2$\rightarrow$3D, $N_\mathrm{int}=10$&
0.27&$2\times 10^{-3}$&0.73
&0.38&0.070&0.12&0.43
&$1.8\times 10^3$
\\\hline

M9&BS in 2$\rightarrow$3D, $N_\mathrm{int}=3$&
0.66&$1\times 10^{-3}$&0.33
&0.67&0.053&0.065&0.21
&$1.3\times 10^3$
\\\hline

M10&BS in 2$\rightarrow$3D, $N_\mathrm{int}=1$&
0.85&$0.02$&0.13
&0.82&0.060&0.022&0.099
&$1.1\times 10^3$
\\\hline

M11&No gas torque on $e$&
0.90&$0.02$&0.08
&0.83&0.064&0.015&0.090
&$1.2\times 10^3$
\\\hline

M12&No gas migration&
0.95&$1\times 10^{-3}$&0.05
&0.92&0.034&0.0099&0.038
&$2.8\times 10^2$
\\\hline

M13&No gas hardening&
0.84&$0.040$&0.12
&0.76&0.071&0.067&0.098
&$1.1\times 10^3$
\\\hline

M14&${\dot M}_\mathrm{out}={\dot M}_\mathrm{Edd}$&
0.86&$0.01$&0.12
&0.78&0.086&0.020&0.12
&$1.1\times 10^3$
\\\hline

M15&$r_\mathrm{out,BH}=0.3\,\mathrm{pc}$&
0.83&0.08&0.08
&0.74&0.10&0.034&0.12
&$4.8\times 10^2$
\\\hline

M16&$M_\mathrm{SMBH}=4\times 10^7\Msun$&
0.87&$0.04$&0.09
&0.78&0.11&0.035&0.076
&$1.1\times 10^3$
\\\hline

M17&$\Gamma_\mathrm{Edd,cir}=10$&
0.92&$0.01$&0.07
&0.86&0.066&0.030&0.048
&$1.1\times 10^{3}$
\\\hline

M18&${\beta}_\mathrm{v}=1$&
0.79&0.15&0.06
&0.70&0.058&0.098&0.15
&$2.9\times 10^2$
\\\hline

M19&$\delta_\mathrm{IMF}=1.7$&
0.92&$7\times 10^{-4}$&0.08
&0.84&0.073&0.018&0.072
&$4.5\times 10^3$
\\\hline

M20&$M_\mathrm{star,3pc}=3\times 10^6\Msun$&
0.90&$0.03$&0.07
&0.81&0.075&0.023&0.094
&$3.5\times 10^2$
\\\hline

M21&$\gamma_{\rho}=1.5$&
0.87&$0.03$&0.10
&0.78&0.072&$0.060$&0.089
&$8.5\times 10^2$
\\\hline

M22& $3\times m_{1g}$
&0.94&$2\times 10^{-3}$&0.06
&0.89&0.051&0.022&0.043
&$1.5\times 10^3$\\\hline

M23&$N_\mathrm{BH,ini}=2000$&
0.85&$0.10$&0.06
&0.69&0.13&0.075&0.10
&$2.2\times 10^{2}$
\\\hline

M24&$N_\mathrm{BH,ini} =2000$, $\beta_\mathrm{v}=1$&
0.28&$0.72$&0
&0.26&0.54&0.065&0.14
&$56$
\\\hline

M25
&$f_\mathrm{pre}=0.7$
&0.91
&$5\times 10^{-3}$
&0.08
&0.87
&0.050
&0.019
&0.065
&$1.3\times 10^3$

\\\hline

M26&No exchange in BS&
0.95&$9\times 10^{-3}$&0.04
&0.88&0.066&0.025&0.028
&$3.7\times 10^3$
\\\hline

\end{tabular}
\end{center}
\end{table*}

\begin{table*}
\begin{center}
\caption{
Main assumptions and simplifications made in our model. 
}
\label{table:assumptions}
\hspace{-5mm}
\begin{tabular}{c|c}
\hline 
1& The AGN disk is in steady state and the SMBH is static\\\hline
2& The formation and evolution of compact objects and their binaries other than BHs are neglected\\\hline
3& Stable triple systems composed of stellar-mass BHs \citep[e.g.][]{Antonini17} are not taken into account \\\hline
4& Massive perturbers \citep{Deme20} are not considered\\\hline
5& In the outer regions of the AGN disk, BHs form directly without experiencing stellar phases \\\hline
6& The AGN disk and the flattened BH component are initially rotating in the same direction\\\hline
7& Gas captured by BHs (binaries) rotates in the same direction to the AGN disk \citep{Lubow99} before viscous torques operate\\\hline
8& The most massive pair remains in a binary after hard binary-single interactions \\\hline
9& Softer binaries are always disrupted after hard binary-binary interactions \\\hline
10& Kick velocities at hard-BS interactions are given by a mode of their distribution referring to \citet{Leigh18} \\\hline
\end{tabular}
\end{center}
\end{table*}

\end{document}